\documentclass[aps,twocolumn]{revtex4}%
\usepackage{amsfonts}
\usepackage{amsmath}
\usepackage{amssymb}
\usepackage{graphicx}%
\begin{document}
\title{Common energy scale for magnetism and superconductivity in cuprates}
\author{Amit Keren and Amit Kanigel}
\affiliation{Physics Department, Technion-Israel Institute of Technology, Haifa 32000, Israel.}
\date{\today}

\begin{abstract}
Many compounds based on CuO$_{2}$ planes (cuprates) superconduct below a
critical temperature $T_{c}$. Some of them show a second phase where a
spontaneous static magnetic field appears below a critical temperature $T_{g}%
$, which is lower than $T_{c}$. By comparing $T_{c}$ and $T_{g}$ in numerous
superconducting families, each with its own maximum $T_{c}$, we find that the
same energy scale determines both critical temperatures. This clearly
indicates that the origin of superconductivity in the cuprates is magnetic. 

\textbf{Comment}: This is an updated version of the original paper published
in Phys. Rev. B \textbf{68}, 012507 (2003) and includes new YBa$_{2}$Cu$_{3}%
$O$_{6+y}$ data from Ref.\cite{SannaPrivate}

\end{abstract}
\maketitle

One of the most challenging tasks of solid-state physics today is to
understand the mechanism for superconductivity in cuprates. These materials,
which have a relatively high critical temperature $T_{c}$, are based on doped
CuO$_{2}$ planes. Since at zero doping they are antiferromagnets, several
theories ascribe their superconductivity to holes interacting via a magnetic
medium \cite{Theory1,Theory2}. Yet the phenomenon of superconductivity begins
at doping levels in which magnetism almost disappears, and therefore there is
no clear evidence relating the two. Fortunately, there is a narrow doping
range in which superconductivity and magnetism, in the form of randomly
oriented static spins (a spin glass), co-exist below a critical temperature
$T_{g}<T_{c}$. We thus focus on this doping range and examine $T_{g}$ and
$T_{c}$ in numerous superconducting families, which are distinct in the sense
that each one has its own maximum $T_{c}$ [$T_{c}^{max}$]. We find that in all
cases a common energy scale controls both critical temperatures. Therefore
magnetism and superconductivity in the cuprates are different facets of the
same Hamiltonian.

\begin{figure}
[h]
\begin{center}
\includegraphics[
natheight=8.661100in,
natwidth=11.087800in,
height=3.0208in,
width=3.64in
]%
{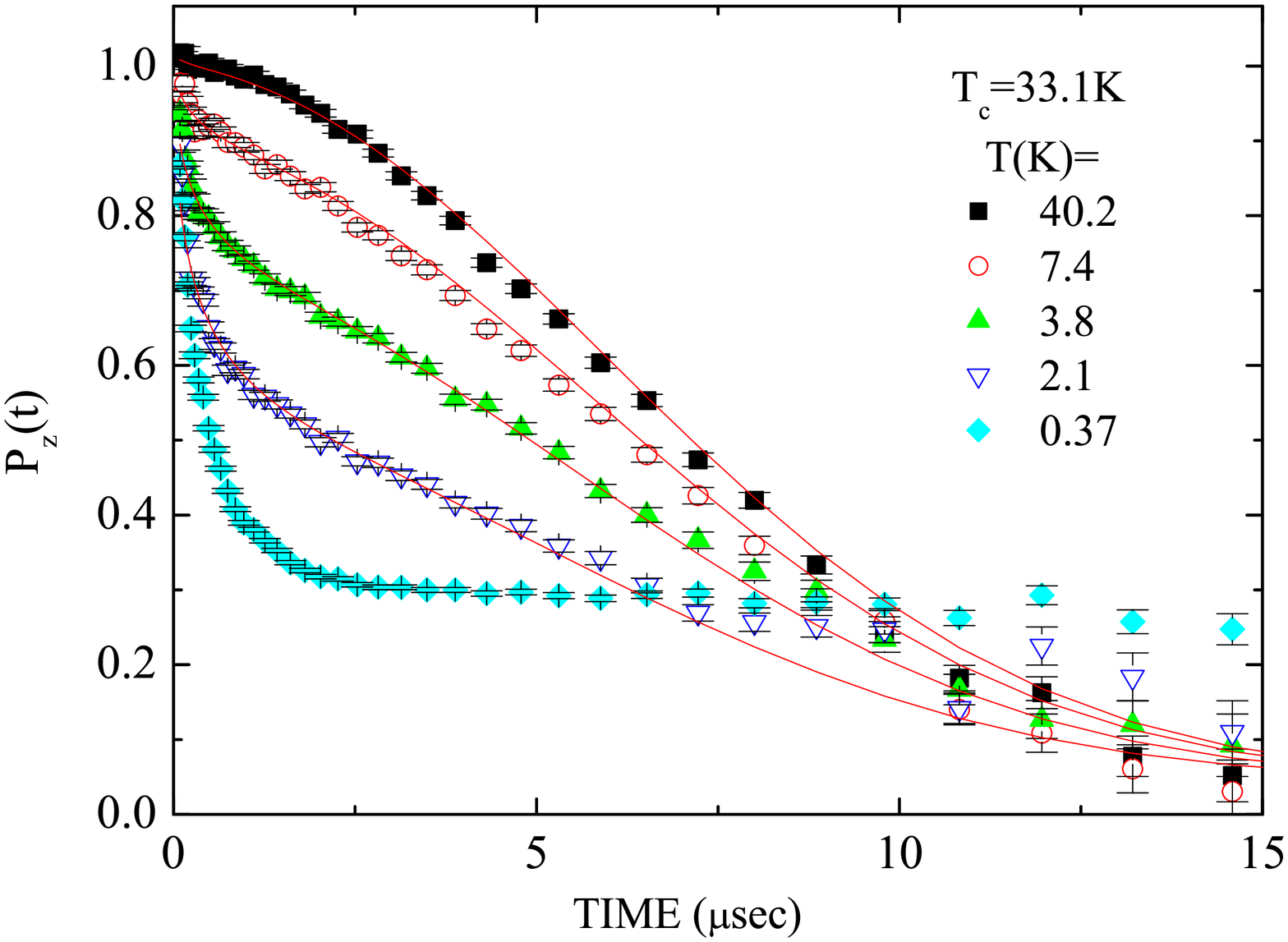}%
\caption{$\mu$SR spectra obtained in a x=0.1, y=7.012 CLBLCO sample at various
temperatures. The solid lines are fits using Eq.~\ref{FitFunc}.}%
\label{Spectra}%
\end{center}
\end{figure}

The families for which both $T_{g}$ and $T_{c}$ data exist are: (Ca$_{x}%
$La$_{1-x}$)(Ba$_{1.75-x}$La$_{0.25+x}$)Cu$_{3}$O$_{6+y}$ [CLBLCO]
\cite{KanigelPRL02}, La$_{2-y}$Sr$_{y}$CuO$_{4}$ [LSCO]
\cite{NiedermayerPRL98,PanaPRB02}, Y$_{1-y}$Ca$_{y}$Ba$_{2}$Cu$_{3}$O$_{6}$
[YCBCO] \cite{NiedermayerPRL98}, Bi$_{2.1}$Sr$_{1.9}$Ca$_{1-x}$Y$_{x}$Cu$_{2}%
$O$_{8+y}$ [Bi-2212] \cite{PanaPRB02}, and YBa$_{2}$Cu$_{3}$O$_{6+y}$ [YBCO]
\cite{SannaPrivate}. Several groups including ours gathered the data, and the
determination of $T_{g}$ was done using the $\mu$SR technique. In this
technique one implants fully polarized positive muons in a sample and measures
the time dependence of their polarization $P_{z}(t)$. This polarization
changes dramatically when static magnetic fields appear. This is demonstrated
for a superconducting compound from the CLBLCO family with $T_{c}=33.1~$K in
Fig.~\ref{Spectra}, which is taken from Ref. \cite{KanigelPRL02} for
completion. Between $T=40$ and $8$~K, $P_{z}(t)$ is typical for muon
polarization in an environment where the magnetic field emanates from nuclear
moments. We denote this polarization by $P_{z}^{\infty}(t)$. At about
$T=7.4~$K a fast relaxation component appears, which is due to some additional
strong magnetic field. As the temperature is lowered the fast relaxing
component grows at the expense of the slow one, and at a temperature of
$0.37~$K, no slow relaxing component is observed. In addition, at this
temperature the polarization saturates at long times at $1/3$ of its initial
value. This is typical for randomly frozen magnetic fields where $1/3$ of the
fields happen to point in the direction of the muon spin.%

\begin{figure}
[h]
\begin{center}
\includegraphics[
natheight=8.313400in,
natwidth=10.683900in,
height=2.7691in,
width=3.5509in
]%
{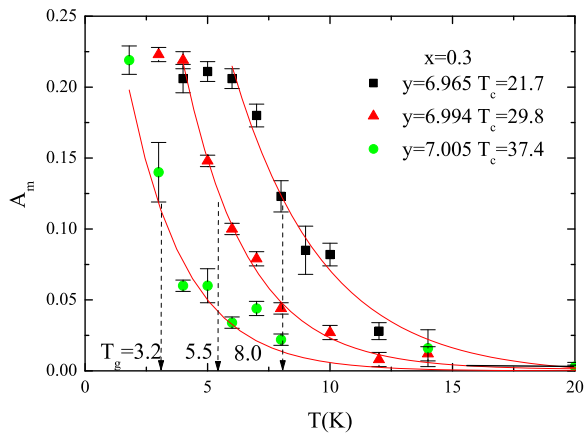}%
\caption{The magnetic amplitude $A_{m}$ as function of temperature for
different (Ca$_{x}$La$_{1-x}$)(Ba$_{1.75-x}$La$_{0.25+x}$)Cu$_{3}$O$_{y}$
samples. The solid lines are guides to the eye. $T_{g}$ is the temperature at
which $A_{m}$ is half of its saturation value, as demonstrated by the dashed
lines.}%
\label{TgDetermination}%
\end{center}
\end{figure}

In order to determine $T_{g}$ quantitatively all authors effectively fit their
data to
\begin{equation}
P_{z}(t)=A_{m}\exp[-(\lambda t)^{\beta}]+A_{n}P_{z}^{\infty}(t)
\label{FitFunc}%
\end{equation}
where $\lambda$ is a relaxation rate, and the amplitudes $A_{m}$ and $A_{n}$
represent muons in magnetic and normal environments. However, different
authors use different parameters in the fit function for the determination of
$T_{g}$. We will show below that this has no bearing on our final conclusion.
In particular, we fit Eq.~\ref{FitFunc} to the data in Fig.~\ref{Spectra} with
$\beta=1/2$ and $A_{m}+A_{n}$ common to all temperatures. In
Fig.~\ref{TgDetermination} we present $A_{m}$ as a function of temperature for
three different samples of the CLBLCO family with x=0.3. As expected $A_{m}$
grows as the temperature decreases and saturates. Our criterion for $T_{g}$ is
the temperature at which $A_{m}$ is half of its saturation value as
demonstrated by the vertical lines. This figure demonstrates the sensitivity
of $T_{g}$ to doping.

In order to quantify the relation between $T_{g}$ and $T_{c}$ we distinguish
between two kinds of holes. The first kind we call \emph{mobile} holes, and
their doping level is $p_{m}$. The second kind is the usual \emph{chemical}
holes and their doping level is denoted by $p$. The reason for this
distinction is that the only experimental known value is that of the chemical
formula of the compounds, namely, the $x$ and $y$ values. Theoretical
arguments relate $x$ and $y$ to $p$
\cite{BrownSSC91,TallonPhysicaC90,ChmaissemPRB01}, but the accuracy of these
relations is debatable \cite{ChmaissemPRB01}. By introducing $p_{m}$ we allow
for an additional scaling parameter, which could be determined experimentally,
and could lead to a comparison between different compounds. An equally good
name for $p_{m}$ could have been \textquotedblleft corrected hole
doping\textquotedblright. The scaling parameter is determined as follows.
First we convert the $T_{g}$ and $T_{c}$ values of all material to be
functions of $p$.%

\begin{figure}
[h]
\begin{center}
\includegraphics[
natheight=10.879300in,
natwidth=8.717300in,
height=4.0274in,
width=3.2335in
]%
{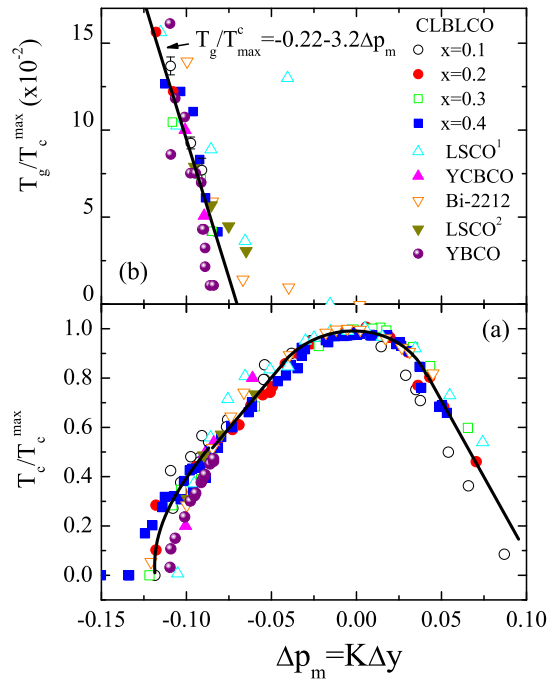}%
\caption{(a) $T_{c}/T_{c}^{max}$ and (b) $T_{g}/T_{c}^{max}$ as a function of
$\Delta p_{m}=K_{f}\Delta p$ (see Eq.~\ref{Holes}). $K_{f}$ is chosen so that
$T_{c}/T_{c}^{max}$ vs. $\Delta p_{m}$ domes of various cuprates families
collapse into a single curve. As a consequence $T_{g}/T_{c}^{max}$ vs. $\Delta
p_{m}$ also collapses into a single line.}%
\label{Scaling}%
\end{center}
\end{figure}

The case of LSCO, YCBCO, Bi-2212, and YBCO is immediate since the authors of
Ref. \cite{NiedermayerPRL98,PanaPRB02,SannaPrivate} present their data in this
way. For CLBLCO however $T_{g}$ and $T_{c}$ are given as a function of $y$
\cite{KanigelPRL02}. We assume the relation $p=-0.205+y/3$ obtained from
simple valance counting. Second, we define $p^{opt}$ as chemical hole doping
at optimum, where optimum means $T_{c}^{\max}$, and introduce $\Delta
p=p-p^{opt}$. Finally we write
\begin{equation}
\Delta p_{m}=K_{f}\Delta p\label{Holes}%
\end{equation}
where $K_{f}$ is the scaling parameter that is different for the various
cuprate families. We interpret $\Delta p_{m}$ as $p_{m}-p_{m}^{opt}$ where
$p_{m}^{opt}$ is the number of mobile holes at optimum. This point requires
extra attention; the scaling we perform between chemical and mobile holes is
done by counting them from optimum, and not from $p=0$. We determine $K_{f}$
from experimental data by making $T_{c}/T_{c}^{max}$, for all the families,
collapse onto one curve resembling the curve of La$_{2-y}$Sr$_{y}$CuO$_{4}$,
since in this case it is believed that $p_{m}=p$. This is demonstrated in
Fig.~\ref{Scaling}a. It should be pointed out that LSCO serves only as a
reference, and whether $p_{m}=p$ for this compound or not has no bearing on
our conclusions. A summary of $p^{opt}$, $K_{f}$ and $T_{c}^{\max}$ is given
in table 1. In Fig.~\ref{Scaling}b we also plot $T_{g}/T_{c}^{max}$ as a
function of $\Delta p_{m}$\ (using the previously determined values of $K_{f}%
$). Magically, $T_{g}/T_{c}^{max}$ also collapse onto one line for all the
cuprates we have examined. The line, depicted in Fig.~\ref{Scaling}b, is
described by
\begin{equation}
T_{g}/T_{c}^{max}=-0.22-3.2\Delta p_{m}.\label{TgoverTcvsDp}%
\end{equation}
Up to date this type of scaling was demonstrated only for the CLBLCO family
\cite{KanigelPRL02}.%

It is important to mention that Eq.~\ref{TgoverTcvsDp} is independent of the
criteria used to determine $T_{g}$. In the case of LSCO, for example, $T_{g}$
was determined from Eq.~\ref{FitFunc} by two different methods. (I) The
temperature at which $\beta=1/2$; a behavior typical of spin glasses at
$T_{g}$ \cite{PanaPRB02}. (II) The temperature where $\lambda$, obtained only
from fit to the long time data with $\beta=1$, has a peak; a common feature of
all magnets upon freezing \cite{NiedermayerPRL98}. Both methods agree with
each other \cite{PanaPRB02}.

We interpret the scaling of Fig.~\ref{Scaling} as follows: The Uemura
relations \cite{UemuraPRL89} and recent theories of hole pair boson motion in
an antiferromagnetic background \cite{Theory2} suggest that $T_{c}$ is
proportional to $n_{s}$ with a proportionality constant $J_{f}$, where the
subscript $f$ stands for family, namely,
\begin{equation}
T_{c}=J_{f}n_{s}(\Delta p_{m}). \label{Uemura}%
\end{equation}
The reason different families have different $T_{c}^{\max}=J_{f}n_{s}(0)$ is
because $J_{f}$ varies from one family to the next, but $n_{s}(\Delta p_{m})$
does not. Therefore,
\begin{equation}
T_{c}/T_{c}^{\max}=n_{s}(\Delta p_{m})/n_{s}(0). \label{Tctons}%
\end{equation}
is a function of $\Delta p_{m}$ for all cuprate families. Using
Eq.~\ref{TgoverTcvsDp} this gives
\[
T_{g}=J_{f}n_{s}(0)\left(  -0.22-3.2\Delta p_{m}\right)  .
\]
Thus, the successes of the simultaneous scaling of $T_{c}$ and $T_{g}$ for all
the compounds discussed here suggests that the same energy scale $J_{f}$
controls both the superconducting and magnetic transitions in all cuprates.

\begin{table}[h]
\centering
\par%
\begin{tabular}
[c]{|l|l|l|l|}\hline
HTSC Familiy & $P_{opt}$ & $K_{f}$ & $T_{c}^{\max}$\\\hline
CLBLCO $x=0.1$ & 0.18 & 2.0 & 58\\\hline
CLBLCO $x=0.2$ & 0.18 & 1.9 & 69\\\hline
CLBLCO $x=0.3$ & 0.18 & 1.8 & 77\\\hline
CLBLCO $x=0.4$ & 0.18 & 1.5 & 80\\\hline
LSCO & 0.16 & 1.0 & 38\\\hline
YCBCO & 0.16 & 1.1 & 65\\\hline
Bi-2212 & 0.16 & 1.1 & 44\\\hline
YBCO & 0.16 & 1.05 & 93\\\hline
LSCZO $x=0.01$ & 0.16 & 1.5 & 26\\\hline
LSCZO $x=0.01$ & 0.18 & 2 & 17\\\hline
\end{tabular}
\caption{Showing the optimal chemical doping, the scaling factor used in
Eq.~\ref{Holes} to produce $\Delta p_{m}$, and the maximum $T_{c}$ for the
varius compounds presented in Fig. \ref{Scaling} The $T_{c}^{\max}$ (and
$p^{opt}$) of YCBCO is not known, and the values given in the table are
assumed. Only two samples of YBCO, for which both $T_{g}$ and $T_{c}$ have
been measured, are shown.}%
\label{Table}%
\end{table}

At first this result seems surprising, since it is believed that in the
antiferromagnetic phase of the cuprates there are three magnetic energy
scales. The isotropic in plane Heisenberg coupling $J$, and the in-plane and
out-of-plane anisotropy energies $J\alpha_{xy}$ and $J\alpha_{\bot}$
respectively. However, Keimer \emph{et al.} showed that the N\'{e}el
temperature $T_{N}$ depends only logaritmically on both anisotropies
$\alpha_{xy}$ and $\alpha_{\bot}$ \cite{KeimerPRB92}. It is conceivable that
this is also the situation in the glassy phase. In that case the energy scale
of $T_{g}$ will be set only by $J$. Another two dimensional theory that
appears to support the existence of glassy freezing is given in
Ref.~\cite{BarneaCondmat03}.

Further insight could be achieved by assuming a linear relation between
$n_{s}$ and $\Delta p_{m}$, namely,
\begin{equation}
n_{s}(\Delta p_{m})=\alpha(p_{m}^{opt}+\Delta p_{m}).\label{phltons}%
\end{equation}
If all the mobile holes had turned into Cooper pairs we would have $\alpha$
$=1/2$ . Taking $p_{m}^{opt}=0.16$, we find from Eqs.~\ref{TgoverTcvsDp} and
\ref{phltons}
\begin{equation}
T_{g}/T_{c}^{\max}=0.3\left[  1-c_{g}\times n_{s}(\Delta p_{m})\right]
\label{TgtoTcmax}%
\end{equation}
where $c_{g}=10.6/\alpha$. This equation could be used to predict $T_{g}$ for
compounds in which the magnetic transition is not found yet.%

\begin{figure}
[h]
\begin{center}
\includegraphics[
natheight=11.143100in,
natwidth=8.633400in,
height=3.0502in,
width=2.3696in
]%
{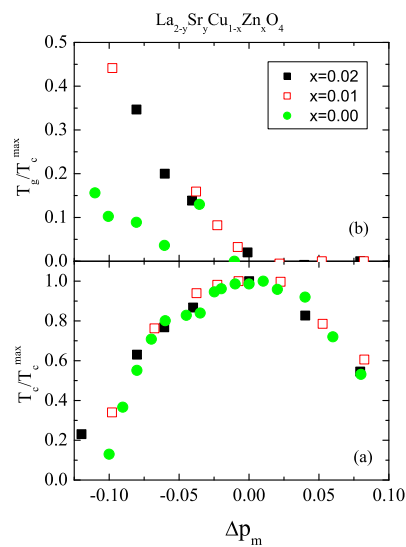}%
\caption{(a) $T_{c}$/$T_{c}^{max}$ and (b) $T_{g}$/$T_{c}^{max}$ as a function
of $\Delta p_{m}=K_{f}\Delta p$ (see Eq.~\ref{Holes}). $K_{f}$ is chosen so
that $T_{c}$/$T_{c}^{max}$ vs. $\Delta p_{m}$ domes for various La$_{2-y}%
$Sr$_{y}$Cu$_{1-x}$Zn$_{x}$O$_{4}$ compounds, representing impure cases,
collapse into a single curve. The same scaling does not apply to $T_{g}$.}%
\label{Znscaling}%
\end{center}
\end{figure}

Finally, it is important to demonstrate that the simultaneous scaling of
$T_{g}$ and $T_{c}$ is a property of clean superconductors and does not work
in all cases. A perfect example for a scaling failure is given by La$_{2-y}%
$Sr$_{y}$Cu$_{1-x}$Zn$_{x}$O$_{4}$ [LSCZO] \cite{PanaPRB02}. Here samples with
the same amount of Zn are considered to be one family of HTSC with its own
$T_{c}^{\max}$. The reduction of $T_{c}^{\max}$ with increasing Zn
concentration is a result of the increasing impurity scattering rates since
the Zn reside in the CuO$_{2}$ plane. As demonstrated in Fig.~\ref{Znscaling},
the scaling transformation that makes all $T_{c}$ vs. $\Delta p_{m}$ domes
collapse into one function does not apply for $T_{g}$ vs. $\Delta p_{m}$. The
parameters used to generate this plot are also given in table 1. The failure
of the scaling suggests that a mechanism with a different energy scale is
involved in the reduction of $T_{c}$ when impurities are present.
Interestingly the two data sets of $T_{g}/T_{c}^{\max}$ vs. $\Delta p_{m}$ for
the impure cases do full on the same line.

We conclude that the variation of $T_{c}$ between different superconducting
families, based on CuO$_{2}$ planes, is a consequence of variations in the
strength of the magnetic interactions.

We would like to thank the PSI and ISIS facilities for their kind hospitality
and continuing support of this project. We acknowledge very helpful
discussions with A. Auerbach and E. Altman. We are grateful for S. Sanna and
R. Derenzi for providing their data prior to publication. The Israeli Science
Foundation and the EU-TMR program funded this work.

\end{document}